# A Distributed Intrusion Detection System Using Cooperating Agents

Jaydip Sen

Innovation Lab, Tata Consultancy Services Ltd
Bengal Intelligent Park, Salt Lake Electronic Complex, Kolkata- 700091, INDIA
{Jaydip.Sen}@tcs.com

**Abstract.** The current intrusion detection systems have a number of problems that limit their configurability, scalability and efficiency. There have been some propositions about distributed architectures based on multiple independent agents working collectively for intrusion detection. However, these distributed intrusion detection systems are not fully distributed as most of them centrally analyze data collected from distributed nodes which may lead to a single point of failure. In this paper, a distributed intrusion detection architecture is presented that is based on autonomous and cooperating agents without any centralized analysis components. The agents cooperate by using a hierarchical communication of interests and data, and the analysis of intrusion data is made by the agents at the lowest level of the hierarchy. This architecture provides significant advantages in scalability, flexibility, extensibility, fault tolerance, and resistance to compromise. A proof-of-concept prototype is developed and experiments have been conducted on it. The results show the effectiveness of the system in detecting intrusive activities.

## 1 Introduction

 A secure computer system provides guarantees regarding the confidentiality, integrity, and availability of its objects (such as data, purpose, or services). However, systems generally contain design and implementation flaws that result in security vulnerabilities. An intrusion can take place when an attacker or a group of attackers exploits the vulnerabilities and thus damages the confidentiality, integrity or availability guarantees of a system. Intrusion Detection Systems (IDSs) detect some set of intrusions and execute some predetermined actions when an intrusion is detected.

Over the last one and half decade, research in the field of intrusion detection has been heading towards a distributed framework of systems that do local detection and provide information to perform global detection of intrusions. These distributed frameworks of intrusion detection have some advantages over single monolithic frameworks. Most of these distributed systems are hierarchical in nature. The local intrusion detection components look for local intrusions and pass their analysis results to the upper levels of the hierarchy. The components at the upper levels analyze the refined data from multiple lower level components and attempt to establish a global view of the system state. However, such IDSs are not fully distributed systems because of the centralized data analysis performed at the higher levels of the hierarchy [1]. In this paper, an agent-based architecture is proposed for performing intrusion detection in a distributed environment. By employing a suitable communication mechanism, the resource overhead is minimized in the distributed intrusion detection process.

The rest of the paper is organized as follows: Section 2 describes some related works done on distributed intrusion detection and their shortcomings, Section 3 briefly introduces the concept of agents and their properties, Section 4 describes the details of the functional architecture of the proposed system, Section 5 gives some details of implementation and experimental results conducted on the system, and finally, Section 6 concludes the paper.

## 2 Related Work

In this section, some of the existing distributed IDS frameworks are discussed briefly. DIDS [2] is a distributed intrusion detection system consisting of host managers and LAN managers doing distributed data monitoring, and sending notable events to the DIDS director. These managers also do some local detection, passing the summaries to the director. The director analyzes the events to determine the security state.

AAFID [3] is a distributed IDS developed in CERIAS at Purdue University. It employs agents at the lowest level of the hierarchy for data collection and analysis and transceivers and monitors at the higher levels for controlling agents and obtaining a global view of activities. It provides a subscription-based service to the agents.

A prototype called the *Hummingbird System* [4] is developed at University of Idaho. It is a distributed system that employs a set of *Hummer* agents, each assigned to a single host or a set of hosts. Each Hummer interacts with other hummers in the system through a *manager*, a *subordinate*, and the *peer* relationships. It enables a system administrator to monitor security threats on multiple computers.

An architecture of an intrusion detection system using a collection of autonomous agents has been proposed in [2]. In cooperation and communication model proposed by the authors, agents request and receive information solely on the basis of their interests. They can specify new interests as a result of a new event or alert. This avoids unnecessary data flow among the agents.

However, most of these intrusion detection systems have the following drawbacks:

(i) *Analysis hierarchy*: as there is a hierarchy in data analysis these systems are very difficult to modify. Changes may have to be made at many levels if any new distributed attack is developed.

(ii) *Data refinement*: when a module from a lower level sends results of analysis to a higher level, some data refinement is done. However, the knowledge of what events are important in a system-wide level is difficult to anticipate at the lower levels of the hierarchy, and thus data refinement may result in loss of important information.

(iii) *Bulky modules*: intrusion analysis engines based on anomaly detection are usually very large modules. They also consume significant amount of system resources in terms of CPU usage, disk I/O, and memory, as they have to analyze long audit trails. In most of the systems described above, these components are present at all levels. They also present multiple points of failures.

(iv) *Passive interactions*: the components of these IDSs interact with each other in a passive way. The lower level components generate data for the upper level components as per the rules driving them. There is no mechanism for a component to query other component on the basis of some analysis that it has done.

To overcome the above problems, an intrusion detection architecture is proposed in this paper that is based on the coordination among a set of agents. The key effort in the proposed approach is directed towards making the agents intelligently cooperate by communicating actively with each other. The intelligence in cooperation is attempted by communicating events and alerts to only those agents that are interested in those events and alerts. The details of the system are discussed in Section 4.

## 3    Agents and Agent-Registries

As discussed Section 1, the proposed system consists of a number of cooperating agents. In this section, the properties of the agents are discussed in brief for an easy understanding of the system architecture.

Crossbie and Spafford [5] first pointed out the suitability of autonomous and intelligent agents in intrusion detection applications. They have defined an agent as a computational or physical entity situated in an environment (either real or virtual), which is able to act in the environment, to perceive and partially represent its environment, and to communicate with other agents. An agent is also driven by internal tendencies (goals, beliefs etc.) and has an autonomous behavior, which is the consequence of its perception, its representation and its interactions with the environment and with other agents. An agent can be described by a set of properties [6]. The ability of an agent to operate without direct intervention of humans or other agents is known as its *autonomy* property. The capability of an agent to integrate itself in a large environment populated by a society of agents with which it has to exchange messages to achieve purposeful actions is called *sociability* property. The ability of an agent to anticipate situations and change its course of action is called *proactivity* property of an agent. *Reactivity* is that property of an agent due to which it reacts in real-time to changes that occur in its environment. *Adaptability* is the ability of an agent to modify its behavior over time to fulfill its problem-solving goals.

*Interests*: an interest is defined as "a specification of data that an agent is interested in, but is not available to it because of the locality of data collection or because it was not primarily intended to observe those data." [2]. Following cases may be considered:

(1) There may be more than one agents residing on the same host that need data from the same data source. Having each agent obtain its data on its own will lead to duplication of effort in terms of accessing the data. Therefore, if the overhead of the data access mechanism from the data source is more than the communication overhead of transferring the data between the agents, it will be more efficient if one agent obtains the data and makes it available to the other agents.

(2) When an agent detects coordinated or distributed attacks, it may need data from agents residing in multiple hosts. The agent may not have access to the data sources in this case and transfer of data from other agents is the only solution. It is impossible for an agent to keep track of all other agents in a large network. Thus an agent may not know the existence or locations of agents those are collecting data in which it may have interest. Hence propagation of interests of different agents in the network becomes essential. The agents may have different types of interest:

*Directed and propagated interests*: an agent knows the host or domain from which it is interested in getting data. If the agent has no specific idea about the source, the interest is propagated across the entire enterprise network. Such an interest is termed as a propagated interest.

(2) *Local-, domain- and enterprise-level interests*: an interest can be a local-level, a domain-level, or an enterprise-level interest depending on whether it is interested in getting data only from a local host, local domain, or the entire enterprise network. The data that is requested by an agent in the form of an interest is delivered to it by the agents(s) servicing that interest. In the proposed scheme, data delivery among agents is suggested using the same hierarchical fashion as in propagation of interests. Direct data delivery without following the hierarchy is not a feasible proposition in a large enterprise network.

*Registries*: as Fig. 1 shows, each host maintains two registries - *Agent Registry* and *Interest Registry*. The agent registry maintains information about the basic agents, the events these agents collect, and the alerts they generate. The interest registry keeps track of the interests originating from the basic agents and interests being serviced.

## 4 Architecture of the System

The proposed architecture is a five-tier hierarchical structure of agents as shown in Fig 2. There are five types of agents are: *local agents* or *basic agents* (BAs), *workstation coordinator agents* (WCAs), *domain coordinator agents* (DCAs), and *security policy manager* agent (SMA). These agents are described below.

### 4.1 Basic (Local) Agents

A workstation (host) can be logically split into multiple entities each performing different functions. Each entity is monitored by a set of BAs. Every BA is responsible for monitoring a part of the system resource, user system, and so on. These agents collect and analyze data flowing into or out of this workstation. Every BA works independently and concurrently with respect to other BAs. The monitoring BA reports the situation detected on an entity, or gives an alarm to the local system user, or takes appropriate measures to prevent the perceived threat, if necessary. This is depicted in Fig. 3. If a BA encounters a situation where it does not have appropriate data to ascertain whether an intrusion is taking place, it expresses an interest in that data and propagates the interest to the WCA on that workstation.

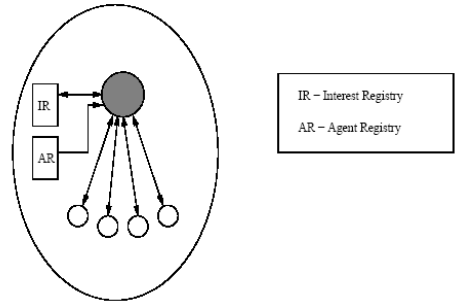

**Fig. 1.** Interest and agent registries

### 4.2 Workstation Coordinator Agents

Every workstation has one WCA. The WCAs serve as links between the BAs and the DCAs. All WCAs residing in the adjacent hosts in the same domain have direct communication links between them. When a WCA receives an interest from a BA, it first checks the type of the interest. If the interest is a host-directed or a domain-directed interest to a different domain, the interest is forwarded to the DCA above the WCA in the agent hierarchy. If the interest is a host-directed interest in the same domain but towards a different host, the interest is forwarded to the appropriate WCA of that host. Otherwise, it refers to the agent registry and determines if there is any agent on this host that can service that interest. If any such agent is found, the WCA informs that agent to send the requested events, and also updates the interest registry in the local host. If no agent on the host is found that can service the interest, and if the interest is a local-level interest, the WCA informs the requesting agent about its inability to service that interest. When a WCA receives an interest from the DCA above or a WCA in the same domain, it looks up the agent registry to determine if there is any agent on the host that can service this interest. If any such agent is found, the WCA requests that agent to send the required data to it, and updates the interest registry. If no such agent is found, the WCA discards the request. When a WCA receives data from a BA in the same host, it looks up the interest registry and sends the data to the BAs whose interest matches the data received. If an agent from a different domain has already registered for the data, the WCA forwards the data to the appropriate DCA for further forwarding. When a WCA receives data from a DCA, it looks up the interest registry, and sends the data to all the BAs in the local host whose interest matches the data received.

### 3.1 Domain Coordinator Agents

Every domain has a DCA. The DCA in each domain has knowledge of all the WCAs in its domain and has direct communication links to each of them. Each DCA is also directly linked to the ECA. The main function of DCAs is to propagate interests and data among WCAs and BAs belonging to separate domains. An interest received by a

DCA from the ECA can only be a propagated interest because any host-directed or domain-directed interests would have been taken care of by appropriate WCA or DCA. The propagated interest received from the ECA is sent to all hosts to which the DCA is connected. An interest received from a DCA of a different domain can be of two types: a host-directed interest to a host in this domain or a domain-directed interest to this domain. The interests of the first type are forwarded to the appropriate hosts, while those of the second type are forwarded to all the hosts in this domain.

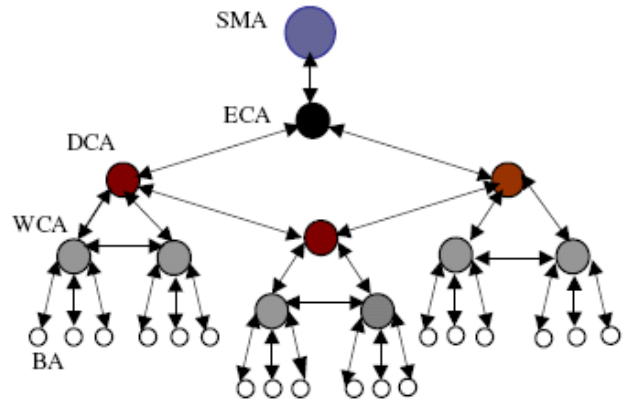

**Fig. 2.** Logical organization of agents. (Bi-directional arrows show flow of data and interest)

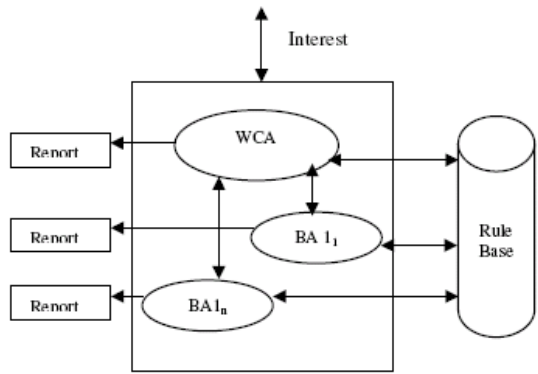

**Fig. 3.** Inter-agent communication in a workstation

An interest received from a WCA in the same domain may be of three types: a host-directed interest to a host in a different domain or a domain-directed interest to a different domain or a propagated interest. The interests of the first and second types are sent to the appropriate DCA for further transfer. The interests of the third type are forwarded to the ECA. An interest directed to a host in the same domain is forwarded to that host. If the directed host is not in the same domain, the DCA forwards it to the ECA. Similarly, a domain-directed interest is sent to all hosts in the domain if the domain specified is its own domain; else it is forwarded to the ECA. Domain-level interests are forwarded to all hosts in its domain. Enterprise-level and propagated interests are not only sent to all hosts in its domain but also forwarded to the ECA.

Data received form the ECA is forwarded only to the host specified. Data received from a host in the domain of the DCA is sent to the DCA of that sent the interest.

### 4.4 Enterprise Coordinator Agent

The ECA is at the top of all DCAs in the agent hierarchy, monitoring all the domains under it. Its functionality is simpler than that of a DCA, since it only has to keep track of the domains under it, and forward data and interests between them.

### 4.5 Security Policy Manager Agent

The SMA manages the security policies specified by the administrator. The administrator interacts with the agents from a high level using security policies that specify the roles and responsibilities of the agents, and the behavior they should exhibit when they receive events or alerts.

Secure and reliable transmission of messages is a prime requirement for operation of a distributed intrusion detection system. In proposed system, two different types of communication are identified: (i) communication among the agents at the same host, (ii) communication among the hosts at different hosts. The important messages exchanged between the agents are: (i) registration request to agent registries, (ii) location queries for agents, (iii) agent queries, and (iv) registry update requests.

For agent communication in the same host, mechanisms like pipes, message queues, shared memory can be used for. In the proposed scheme, the shared memory mechanism is used since it is efficient for one-to-many communication [3]. For agents in different hosts, the *agent management system* (AMS) of JADE [7] has been used. This reduces the cost of communication, as the methods of communication are not replicated for each agent. A PKI model is implemented for providing two-way authentication of messages and the agents.

## 5   Implementation and Experiments

In this section, some implementation details and experimental results are presented.

A proof-of-concept prototype has been built with JADE [7] and Java. JADE (Java Agent Development Environment) is a middleware developed by TILAB that enables faster development of multi-agent distributed applications based on the peer-to-peer communication architecture. JADE has been implemented fully in Java. It includes both the libraries (i.e. the Java classes) required to develop application agents, and the run-time environment that provides the basic services and that must be active on the host before agents can be executed. From the functional point of view, JADE provides all the basic services necessary for distributed peer-to-peer applications. It allows each agent to dynamically discover other agents and to communicate with them by message passing mechanism. The agents communicate by exchanging asynchronous messages. Each agent is identified by a unique identifier and provides a set of services. An agent can register its services and search for other agents.

Implementation of the agents under JADE framework involves the following steps:

1. Determination of the agent behavior.
2. Implementation of the agent class (extending the existing classes of JADE).
3. Implementation of agent meta-behavior by instantiating an existing class or introducing a new class and then instantiating it. The meta-behavior provides an agent with a self-control mechanism to dynamically schedule its behavior in accordance with its internal state.
4. Instantiation of the agent class.
5. Initializing the agent acquaintances.
6. Deployment and activation of the agents.

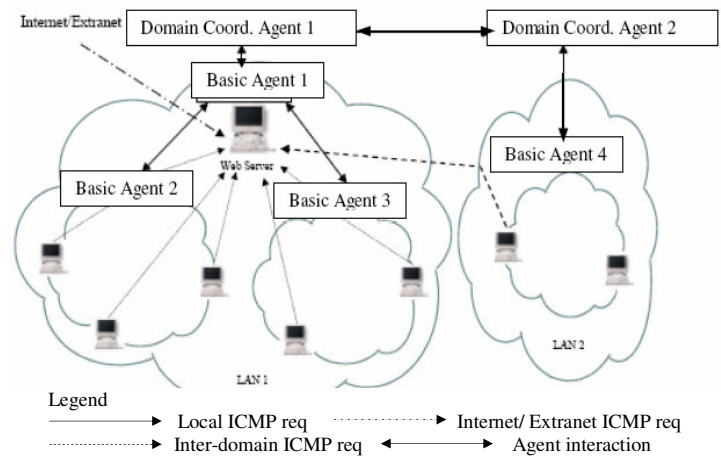

**Fig. 4.** An illustration of the detection of an ICMP-flooding attack

To illustatrate the cooperative framework, a scenario is described where an ICMP flooding attack on a web server is detected (Figure 4). In the web server of a LAN 1 (domain 1), the basic agent 1 monitors all the ICMP requests entering the web server. These requests may come from either the local hosts of the same LAN1 (domain 1), or local hosts from different a domain, such as LAN2, or from external hosts anywhere in the Internet/Extranet. The basic agent 1 of the web server plays the most important role here, because its goal is to observe the number of ICMP requests received by the web server over a certain period of time, and block the access to the server if it finds that the number of requests exceeds a certain predetermined threshold specified in its rule of detection. For this purpose, the basic agent 1 collects all ICMP request data that arrives at the web server. It also interacts with basic agent 2 and basic agent 3 on some hosts located in the same domain as the web server to obtain their collected data (for instance, number of ping probes they collect from the hosts they supervise and send to the web server). The same interest request and data delivery can take place between the basic agent 1 and another basic agent (e.g., basic agent 4) on a host belonging to a different domain through their domain coordination agents. These events are used to update/create the beliefs of the basic agent 1. If the goal is reached, an alarm is raised.

Experiments have been conducted to test the performance of the developed prototype. In the experiments, the KDD Cup 1999 intrusion detection contest data [8] have been used. This data was compiled by the 1998 DARPA intrusion detection and evaluation program by MIT Lincoln Lab. The original database consisted of 4.94 million records. There were 41 attributes for each record. Each record was also given a class label that specified the category of attack to which the record belonged. All the attacks were categorized in to four major categories: i) Denial of Service (DoS), ii) Remote to User (R2C), iii) User to Root (U2R), and iv) Probing (Probe).

A dataset is constructed that consists of 15000 records by randomly selecting records from the original database, such that the number of data instances selected from each class was proportional to their frequency in the original database. One additional class of records is added what is called as the 'normal' class in addition to the 4 attack types mentioned above. The knowledge about these attacks is then distributed among the agents of the system.

To test the performance of the the of the prototype system that we developed we conducted experiments in a network of workstations each having Pentium 4 processor with 3GHz clock speed and 1 GB RAM. The data rate of the Ethernet was 100 Mbps. The agents were installed in the workstations in a hierarchical fashion.

Using *Ethereal* network sniffer, the network is monitored and the bandwidth utilization of the agents is evaluated. First, the network is monitored with no agents running on any workstation so as to establish the baseline utilization. The agents are then activated on each workstation and simulated attacks are launched from different workstations into the network. From the data collected by the sniffer, it was evident that the agents had very little bandwidth consumption. During the one-hour time when the agents were active on the workstations, only 15% increase in number of packets captured in the network by the sniffer was compared with the one-hour period when the network was monitored without activating the agents. The average bandwidth consumption by the agents never exceeded 5% of the 100 MBPS Ethernet.

To test the CPU utilization by the agents, some workstations are randomly selected and some intensive user programs are activated on them. The IDS agents are then invoked on these workstations, and some elementary attacks like password guessing are simulated. During the thirty-minute analysis period, the maximum CPU utilization of the agents was found to be only 8.76%, the average utilization being 0.35%.

Finally, for testing the attack detection capability and false positive rates, 37 different attacks are simulated from some workstations on the others. Some of attacks chosen those are not in the knowledge base of the agents. This is done to test the ability of the IDS prototype to detect novel attacks. Table 1 summarizes the results of the experiment. The results show that the performance of the prototype is quite encouraging particularly in terms of successful detection of attacks.

**Table 1.** Performance results of the prototype IDS

| Activity Type | Detection Rate | False Positive Rate |
|---|---|---|
| Denial of Service (DoS) | 93.28% | 11.42% |
| Remote to User (R2U) | 89.41% | 16.53% |
| User to Root (U2R) | 85.63% | 18.34% |
| Probe | 92.26% | 14.38% |
| Normal | 96.30% | 8.24% |

## 7 Conclusion

In this paper, we have discussed some existing intrusion detection systems and illustrated their limitations. Flexibility, autonomy, adaptability, and distribution are the principal features to be addressed to build a suitable architecture that can be useful for detection of complex and distributed attacks. A prototype architecture of an intrusion detection system is presented, which is based on a group of agents that do data collection and analysis in a truly distributed manner. These interest-driven cooperating agents make the system more fault-tolerant. Moreover, the autonomy given to the agents makes the administration tasks of the security officer much easier. Experiments conducted on the system have produced results that show effectiveness of the system.